\documentclass[preprint2]{aastex}

\begin{document}
\title{Advection-Dominated Accretion with Infall and Outflows}
\author{Thomas Beckert \altaffilmark{1,2}}
\altaffiltext{1}{Harvard-Smithsonian Center for Astrophysics, Cambridge,
                 MA 02138, USA}
\altaffiltext{2}{Max-Planck-Institut f\"ur Radioastronomie,
                 Auf dem H\"ugel 69, 53121 Bonn, Germany}

\begin{abstract}
 We present self-similar solutions for advection-dominated accretion
flows with radial viscous force in the presence of outflows
from the accretion flow or infall. The axisymmetric flow is treated
in variables integrated over polar sections and the effects of infall
and outflows on the accretion flow are parametrised for possible
configurations compatible with the self-similar solution.
We investigate the resulting accretion flows for three different
viscosity laws and derive upper limits on the viscosity parameter $\alpha$.
In addition, we find a natural connection to non-rotating and
spherical accretion with turbulent viscosity, which is assumed to persist
even without differential rotation. Positive Bernoulli numbers for
advection-dominated accretion allow a fraction of the gas to be expelled
in an outflow and the upper limit on the viscosity predicts that outflows 
are inevitable for equations of state close to an ideal gas. 
\end{abstract}


\section{Introduction}
Advection-dominated accretion flows (ADAFs) have been invented to
explain low-luminosity black hole candidates like Sgr A$^*$ \citep{na98} in
our Galactic Center. The low-luminosity of this model is achieved in 
an optically thin plasma, where most of the energy is stored in hot ions,
while electrons as potential radiators are inefficiently coupled to the
heat source and remain relatively cold. The electrons become nonetheless
mildly relativistic close to the central black hole and the inevitable
synchrotron radiation is observed from most ADAF candidates. The presence
of magnetic fields not far from energy equipartition with the gas is
indicative of their origin in MHD-instabilities \citep{bal91} leading
to turbulence in the accretion flow and subsequent generation of an
effective viscosity. On larger scales magnetic fields are likely to be
responsible for the collimation of outflows from accretion disks into jets,
seen in the cores of  M87 \citep{rey96} and NGC 4258 \citep{lasota96,her98}, 
which are prototypical ADAF candidates. Furthermore the model can explain
the accretion in some low-luminosity AGNs of elliptical galaxies
\citep{dim99} and in  NGC 4258 \citep{gam99}. In a recent X-ray survey
\citep{sam99} a few more examples have been found for low-luminosity core
in radio-loud AGNs, which are candidates for advection-dominated
accretion flows in their central engines, suggesting that ADAFs can be
found even in radio-loud AGNs and that jet formation is a common feature.
        
Outflow models for ADAFs have been investigated by \citet{blan99} and 
applied to several candidates \citep{dim99, qua99}. On the theoretical
side \citet{igu99} and \citet{stone} have performed time-dependent 2D
calculations of accretion flows, which in some cases resemble ADAFs
for certain viscosity parameters $\alpha \approx 0.1$, but suggest the
production of outflows for larger $\alpha$ \citep{igu99}. It is found
that the $(rr)$ stress tensor component, which was not included
in the original description of vertically integrated models for
accretions flows, is important in the cited calculations for ADAFs.
The existence of self-similar solutions with a radial viscous force
has been shown in previous work \citep{ny95} and discussed for 2D
solutions with a separation of variables.

In this paper we describe advection-dominated accretion flows in
polar-integrated variables including the radial viscous braking force,
which either produce outflows or are formed by wind infall. The wind
infall is assumed to consist of free falling low angular momentum gas,
which is cold with respect to the already existing accretion flow.
One possible source for this gas are stellar winds from massive stars
in young clusters as in the center of our galaxy. In the following we
will refer to this wind infall into the accretion flow as infall.
In our treatment the main difference between infall and outflow is
the sign of the mass infall rate and we talk about winds, if it is not
necessary to distinguish between infall and outflow. We restrict
the discussion to an extension of the self-similar solutions
given by \citet{ny94} for the Newtonian limit.  

In \S\ref{WO} we present the equations, which describe the accretion flow
including the reaction to winds. The role of $(rr)$ stresses and
bulk viscosity is emphasised. We discuss angular momentum transport
and viscosity in \S\ref{Vis} and specify the possible equations of state 
and the resulting energy equation in \S\ref{equStat}. General features
of self-similar solutions are presented in \S\ref{SelfS} and 
in \S\ref{alphaD} detailed solutions for the $\alpha$-viscosity law are
shown. Consequences of the alternative $\beta$-viscosity are discussed 
in \S\ref{BetaDisk} and ADAFs with an intermediate shear-limited 
viscosity law follow in \S\ref{shearADAF}. We compare the solutions and
draw our conclusions in \S\ref{Conclud}. 
%
%-----------------------------------------------------------------------
\section{Stationary accretion with infall or outflows}  \label{WO}
In a first step we have to establish the set of equations, which describe
the accretion flow. Advection-dominated flows, in which  we are
interested, are known to be quasi-spherical \citep{ny95} and therefore
we will use spherical coordinates in our discussion. Consider a stationary,
axisymmetric and rotating flow with angular velocity $\Omega$
around a compact object of mass $M$. Instabilities in the flow, either
hydrodynamic or magneto-hydrodynamic in origin, generate turbulence on 
small scales and lead to an effective viscosity much larger than the
microscopic one.  The  effective turbulent viscosity $\nu$ in the flow
will redistribute specific angular momentum $\ell(r,\theta) =
\sin^2\theta r^2 \Omega$ by a local viscous torque between neighbouring
rings or shells. Short term evolution on scales smaller than the
mean free path of eddies $\lambda$ in the turbulent flow,
which is related to the viscosity $\nu = v_{\rm eddy} \lambda$ is not
resolved in this description. All quantities like density or accretion
velocity must be understood in a local time averaged sense with
probably large short term variations. Besides the time average,
we will discuss accretion flows also as polar-averaged, one-dimensional
flows with only a radial coordinate $r$. The average is taken over
sections of shells occupied by the flow.
%--------------------------------------
\begin{figure}[t]
  \plotone{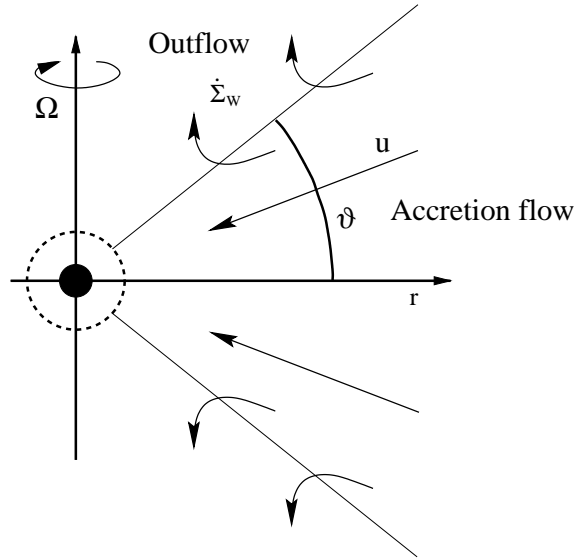} 
\figcaption[fig0.eps]{\label{sketch}
  Schematic view of accretion flow and associated outflow. The
  relativistic region around the central black hole (dashed line) is
  excluded from consideration. The axisymmetric flow is rotating with
  angular velocity $\Omega$ and has a large accretion velocity $|u|$.
  The flow has an opening angel $\vartheta$ and polar-averaged
  quantities are averaged between $-\vartheta$ and $\vartheta$ about
  the equator. The accretion flow may loose mass to an outflow
  as shown or acquire mass with an infall rate $\dot{\Sigma}_W$.}
\end{figure}
%--------------------------------------
Polar motions can consequently
not be traced in this treatment. We are left with the angular velocity
$\Omega=\Omega(r)$ and the radial component of the velocity, which is
the accretion velocity $u=u(r)$. Mass is accreted by the central object
with a rate $\dot{M} = - 2\pi r u \Sigma$. In this 
discussion $\Sigma$ is a suitable polar integral of the density 
%----------------------------------
\begin{equation} \label{SigmaF}
 \Sigma = 2\int_{\pi/2- \vartheta}^{\pi/2+ \vartheta} {\rm d} \theta\, 
 r \sin\theta \rho
\end{equation}  
%----------------------------------
so that $H = r \vartheta$ is the vertical thickness of the accretion flow 
and the integral in equation (\ref{SigmaF}) is restricted to a region
around the symmetry plane of the flow to make room for winds seen
in Fig.\ref{sketch}.  Nonetheless any vertical motion has a radial
and polar component and the later is considered as a source or sink
of mass, momentum and energy. We regard these sources as external to
our description of the flow and call it infall into the accretion flow
or an outflow respectively. 
 
For a non-relativistic flow the conservation of mass implies that the
change in mass flux at every radius is balanced by mass exchange with
the wind in the stationary case
\begin{equation} 
  \frac{1}{r} \frac{\partial (r u \Sigma)}
  {\partial r} =  \dot{\Sigma}_W \quad ,
\end{equation}
where $\dot{\Sigma}_W$ is the mass per surface area added to the 
flow from the infall per time at a given radius. According to equation
(\ref{SigmaF}) this happens at a height $r \vartheta$, but rapid
mixing with the flow is assumed so that the equations averaged over
polar section occupied by the accretion flow remain valid. 
Angular momentum is a conserved quantity in accretion flows without
winds and is redistributed by torques generated by viscous stresses
$t_{r\phi}$ 
\begin{equation} \label{Angu}
  \frac{1}{r} \frac{\partial}{\partial r}(r u \Sigma \ell - r^2 t_{r\phi}) 
  = \dot{\Sigma}_W\ell_W \qquad .
\end{equation}
Here $\ell$ is a mass weighted polar average of $\ell(r,\theta)$.
The infall contributes its own angular momentum to the flow
$\dot{\Sigma}_W\ell_W$ and exerts an additional torque, which depends 
on the relative angular momentum of infall and flow $\ell_W - \ell$.

The accretion velocity $u$ follows from the imbalance of gravitational
force, centrifugal barrier and radial pressure gradient. In addition
to that we include the viscous braking force $F_\nu$
from the $(rr)$-component of the stress tensor and bulk viscosity 
\begin{eqnarray} \label{AcVelo}
  \frac{1}{r}\frac{\partial (r u \Sigma u)}{\partial r} & = &  
  \Sigma \left(r \Omega^2
  - \frac{G M}{r^2}\right)  
  -  r\frac{\partial {\cal P}}{\partial r} \nonumber  \\ & + &
   F_\nu + \dot{\Sigma}_W u_W
\end{eqnarray}
with the integrated pressure ${\cal P} = \int {\rm d} \theta P$ and
the radial velocity of the wind $u_W$, which adds its momentum to
the flow. The change in specific radial momentum induced by winds is
proportional to the velocity difference $u_W - u$.  The viscous force 
(see Appendix \ref{VisF})
\begin{eqnarray}
  F_\nu & = & F_{rr} + F_{\rm bulk} \\ \label{rrshear}
 F_{rr} & = & \eta_1\left[\frac{4r}{3}
 \frac{\partial}{\partial r} 
 \left(\nu\Sigma\frac{\partial}{\partial r} \left(\frac{u}{r}\right)\right) 
  \right. \nonumber \\ & + & \left.
  4 \nu\frac{\partial}{\partial r} \left(\frac{u}{r}\right)\right] \\
   \label{Fbulk}
F_{\rm bulk} & = & \eta\,r\frac{\partial}{\partial r} \left( 
  \frac{\nu\Sigma}{r^3}
   \frac{\partial (r^2 u)}{\partial r} \right)
\end{eqnarray}
has a contribution proportional the $(rr)$ component of the shear tensor 
of the flow in equation (\ref{rrshear}) and the compression of the gas in
equation (\ref{Fbulk}). We make the crude assumption that the
corresponding viscosities are the same for the $(r\phi)$ and $(rr)$
components of the shear stress and also for the bulk viscosity. 
Our ignorance is cast into the parameters $\eta_1$ and $\eta$,
which measure the strength of the $(rr)$-shear and bulk viscosity
relative to the familiar $(r\phi)$-shear viscosity respectively.
For isotropic turbulence $\eta_1 \approx 1$ is expected, but no
estimate on $\eta$ can be derived from this assumption. We will call
$\eta$ and $\eta_1$ viscous force measures in the following.
Finally an energy equation must be given 
\begin{eqnarray} \label{Energ}
  u \Sigma \frac{\partial e}{\partial r} & = & 
  \frac{\Sigma}{r^2}\frac{\partial (r^2 u)}{\partial r} 
  + Q^+ - \Lambda \nonumber \\ & + & \dot{\Sigma}_W\left(\omega_W - \omega
  + \frac{({\bf v_W - v})^2}{2}\right) \quad ,
\end{eqnarray}
which describes the change of specific internal energy $e$ of the
flow due to compression (first term right side), viscous heating $Q^+$,
radiative cooling $\Lambda$, the enthalpy difference between flow
$\omega$ and wind $\omega_W$ and the kinetic energy associated with
the velocity difference, which has to be dissipated into heat.
Here ${\bf v}$ is the velocity vector of the flow and ${\bf v_W}$
the corresponding vector for the wind. 
%
%-------------------------------------------------------------
\section{Angular Momentum Transport and Viscosity}\label{Vis}
The $(r\phi)$ component of the stress tensor is assumed to be proportional
to the corresponding component of the shear tensor with the kinematic 
viscosity $\mu  = \nu \rho$ as the factor of proportionality. 
For the viscosity $\nu$ we will adapt three different representations.
The first and obvious one is the $\alpha$-prescription introduced
by \citet{sun73} and the resulting $\alpha$-disks will be discussed
in \S\ref{alphaD}. The $\alpha$-viscosity is not a unique choice and
effects of the so-called $\beta$-viscosity introduced by Duschl,
Strittmatter \& Biermann are discussed as $\beta$-disks in
\S(\ref{BetaDisk}). For the $\alpha$-viscosity  
\begin{equation} \label{alphvis}
  \nu = \alpha \frac{c_s^2}{\Omega_K}
\end{equation}
we avoided to introduce the vertical scale-height $H$ of the disk,
because we believe that it leads to the misunderstanding that angular
momentum transport in the radial direction depends on the thickness
of the disk. Only with the assumption of vertical hydrostatic
equilibrium is it possible to introduce $H=c_s/\Omega_K$ in
(\ref{alphvis}).

The $\alpha$-viscosity can be recovered from the $\beta$-viscosity law 
in the case of shock-limited turbulence in a Keplerian disk.
The $\beta$-viscosity assumes that a typical length scale in the
direction of transport $\Delta r$ and the typical velocity difference
$\Delta v_\phi$ between interacting shells, which exchange eddies,
equal the mean free path of eddies and their typical velocities
relative to the mean flow respectively. A parametrisation with a new
constant $\beta$ is suggested 
\begin{equation} \label{BetaV}
  \nu = \Delta v_\phi \Delta r = \beta v_\phi r \quad .
\end{equation}
If the typical velocity of eddies in a differentially rotating disk 
is limited by the sound speed, than the typical length scale $\Delta r$
of communication between differentially rotating rings is estimated 
Duschl et al. to be
\begin{equation} \label{lenghtnu}  
  \Delta r \le \frac{c_s}{\Omega} \quad .
\end{equation}
Only in disks with Keplerian rotation does this length scale equal the
vertical scale-height and the form $\nu = \alpha c_s H$ is recovered. 
As a third possibility we consider the shear-limited form of the 
$\beta$-viscosity with the length scale from equation (\ref{lenghtnu}),
which has not been used for sub-keplerian accretion flows before. This is
reasonable, if the mean free path of eddies is not limited by the vertical 
scale-height, but instead determined from the distance between shells, for
which the velocity difference due to differential rotation equals the 
eddie velocity. This argument implies a larger effective viscosity for
disks in sub-keplerian rotation. The $(r\phi)$ component of the viscous
stress tensor becomes in any case
\begin{equation}
  t_{r\phi} = \nu \Sigma r \frac{\partial \Omega}{\partial r}\quad .
\end{equation}
We assume that the same viscosity prescription can be applied to the
$(rr)$-component of the stress and the bulk viscosity. Their relative
strength is scaled to the $(r\phi)$-viscosity by the force measures
$\eta_1$ and $\eta$ introduced above. The heat generated by the
described viscous forces has to be included in the energy equation
and amounts to
\begin{eqnarray} \label{heatQ}
  Q^+ &  = &  \nu \Sigma \left(r \frac{\partial \Omega}{\partial r}\right)^2
  + \frac{4}{3} \eta_1 \nu \Sigma  
  \left( r \frac{\partial}{\partial r}
  \left(\frac{u}{r}\right)\right)^2 \nonumber \\ & + &
   \eta\frac{\nu \Sigma}{r^4}
   \left(\frac{\partial (r^2 u)}{\partial r}\right)^2  \quad .
\end{eqnarray}
The internal energy is increased by viscous heat and compression and
reduced by radiative cooling $\Lambda$.
%
%-------------------------------------------------------
\section{Equation of State}\label{equStat}
In general accretion flows are expected to  exist for all reasonable
equations of state for the accreted gas. The self-similar solutions
we will use in what follows, are restricted to rather simple equations
of state. For ADAFs the gas should be well described by an ideal gas
law with $\gamma = 5/3$ for ratio of specific heats, because the
temperatures in ADAF solutions are so large ($T > 10^7$K for most radii
of interest) that the gas is completely ionised and only ions of a few
heavy elements retain their highly bound inner electrons
(see \citet{nar99} for possible emission lines of these elements).
The change of the equation of state due to partial ionization can be
neglected, but the contribution of magnetic to the total pressure must
be considered for two reasons. The first is the observational evidence
that at least some candidates (e.g. Sgr A$^*$, see \citet{na98}) for
ADAFs show a significant contribution of synchrotron emission to the
total luminosity, which requires magnetic fields close to pressure
equipartition with the gas. Given their existence, we might try a
MHD-description of the accretion flow to separate the evolution of
magnetic fields and gas, which might be different \citep{bis97} unless
magnetic diffusivity \citep{hey96} and/or reconnection require an even
more complex model. The easy way out is to assume that turbulence
generates small scale magnetic fields, which dominate the energy density
in magnetic fields and produce an isotropic contribution to total pressure
and energy. In doing so, one arrives at a hydrodynamic description
of the accretion flow with a equation of state, which has to incorporate
the magnetic pressure. Consider an equation of state of the quite
general form
\begin{equation} \label{EqState}
  P \sim \rho^{\chi_\rho} T^{\chi_T}
\end{equation}
with the exponents $\chi_\rho, \chi_T$ defined according to \citet{cox68}.
For the self-similar solutions we have to restrict the internal energy
per unit volume to be proportional to the total pressure 
\begin{equation} \label{strictE}
 P = \frac{\chi_\rho}{\chi_T}(\gamma -1)e = (\Gamma_3 -1)e
\end{equation}
with $\gamma = c_P/c_V$ being the ratio of specific heats and $\Gamma_3$
the adiabatic coefficient defined by \citet{chand39}. The restriction
from equation (\ref{strictE}) is that $\chi_\rho, \chi_T$ and $\Gamma_3$ 
have to be constants, which requires that the relative strength of
the magnetic pressure must be constant so that the effective equation
of state is an ideal gas law with a ratio of specific heats less
than $5/3$. We define the isothermal sound speed
$c_s = \sqrt{r {\cal P}/\Sigma}$ from the integrated pressure and density. 
The sound speed used in the viscosity prescription is the total 
pressure divided by the mass density. We use thermodynamic relations
\citep{cox68} to get an equation for the temperature from
equation (\ref{Energ}) for stationary accretion flows
\begin{eqnarray} \label{Temp}
  u c_V\frac{ \partial T}{\partial r} & = & \frac{\chi_\rho
  r c_s^2}{\Sigma}
  u \frac{\partial}{\partial r} \left(\frac{\Sigma}{r}\right) 
  + \frac{Q^+ - \Lambda}{\Sigma} \nonumber \\ & + &   
  \frac{\dot{\Sigma}_W}{\Sigma}\left(\omega_W - \omega + \frac{({\bf v_W -
  v})^2}{2}\right) \quad .
\end{eqnarray}
Here $c_V$ is the specific heat at constant volume 
\begin{equation}
  c_V = \frac{\chi_T^2 c_s^2}{\chi_\rho (\gamma -1) T}
\end{equation}
and if we assume that gas pressure contributes a fraction $\delta$ to the
total pressure and magnetic fields are responsible for the rest, we get
\begin{equation} \label{Ga3}
 \Gamma_3-1 = \frac{\delta}{\frac{3}{2}\delta+ (1-\delta)}
\end{equation}
for the adiabatic exponent $\Gamma_3$ which is related to $\gamma$ by
equation (\ref{strictE}). For constant $\chi_\rho$ and $\chi_T$ we can
rewrite the temperature gradient in equation (\ref{Temp}) as a gradient
of the sound speed and search for solutions in terms of the sound speed.
%
%------------------------------------------------------------
\section{Self-similar ADAF solutions with a wind}\label{SelfS}
The set of equations described in \S\ref{WO} and \S\ref{Vis} allow
self-similar power-law solutions for accretion velocity, angular velocity
and sound speed in the way \citet{ny94} have shown
\begin{equation}
  u = - u_0 s^{-1/2} c \qquad \Omega = \Omega_0
  s^{-3/2} \frac{c}{r_G}
\end{equation}
\begin{equation}
   c_s^2 =  a^2 s^{-1} c^2 
\end{equation}
with the radial coordinate scaled to the gravitational radius of the
central mass 
\begin{equation} 
  s = \frac{r}{r_G} \quad; \qquad r_G = \frac{G M}{c^2}
\end{equation}
and the speed of light $c$. It is required that the cooling is either
completely negligible or is a radius independent fraction of the heating
rate 
\begin{equation}
   f = 1-\frac{\Lambda}{Q^+} 
\end{equation}
so that $1-f$ is the cooling efficiency, which will be small, if
electrons are inefficiently coupled to the heat source. This happens, if
ions are preferentially heated by viscous friction and the energy transfer
rate between electrons and ions is smaller than the heating rate.  

For the self-similar solution the radial component of the wind velocity
must be proportional to the accretion velocity and we get a measure
$\xi_1 = u_W/u$, which tells us the relative radial velocity of the wind,
but we will neglect it in the following discussion and assume $\xi_1 = 0$.
This is motivated from the infall calculations of Coker, Melia \& Falcke
(1999), which predicts a very small radial velocity of  the infalling
material in the disk mid-plane for the infall from stellar winds in the
Galactic Center. If a fraction $f$ larger than a few per cent of the
viscously generated heat is not radiated away but kept in the flow as
internal energy, the disk scale height $H = c_s/\Omega_K$
is of the same order as the radius and the infall may have a significant 
radial velocity component. Nonetheless we will neglect it in the following
discussion. The same is true for the enthalpy of the infalling material, 
but here we have better reasons to believe that the gas joining the ADAF
is cold compared with the gas in the ADAF, which is heated by viscous
friction, while the infalling gas maybe provided by stellar winds or
the gas of the H{\sc ii} region Sgr A West in case of the Galactic Center 
with temperatures of  $6\,10^3 - 10^4$ K (see \citet{mez96}
for a review of the Galactic Center). This should be compared to typical
temperatures of ADAFs of $10^7$ K at $10^4$ black hole radii. The 
situation with outflows might be different, where it is hard to imagine
that the gas leaving the disk has a different temperature than the gas
left in the flow. The gas might need some outward pointing radial
momentum to get away from the ADAF, but again we will ignore it here. 
For infalling material we expect it to be in free fall so that their
total velocity is $v_W = v_{\rm ff} = \sqrt{2} c s^{-1/2}$.
In the case of outflows we require that they are able to escape the
gravitational attraction of the central object and their velocity
must be at least the free fall velocity $v_{\rm ff}$.     

In the same way the rotation of the infall has to be a constant
fraction $\xi = \Omega_W/\Omega$ of the rotation of the accretion 
flow itself and again calculations by \citet{cok99} for the Galactic
Center predict infall with only weak rotation. Outflows on the one hand
might be expected to carry their initial angular momentum from the point
of origin and therefore $\xi = 1$. On the other hand magnetically driven
outflows will exert a torque on the remaining accretion flow if the gas
in the outflow has different angular velocity than the point in the flow,
to which it is connected by magnetic field lines.
In this case $0<\xi<1$ would be expected. 

The rate of mass added by the infall per surface area of the flow in 
its central plane is constrained to steep radial profiles and 
introduces one free parameter $p$ (Blandford \& Begelman, 1999)
not to be confused with the pressure $P$
%----------------
\begin{equation}\label{Masswind}
 \dot{\Sigma}_W = -p\frac{ u \Sigma}{r}\quad , \qquad \Sigma =
 \Sigma_0 s^{-1/2 + p} \quad .
\end{equation}
%----------------
For infall into accretion flows $p$ has to be negative and the radial
dependence of the infall rate must be steeper than $r^{-2}$. Therefore
most of the mass is added at small radii and the total mass infall 
$d\dot{M}_W \sim dr\ r^{-1+p}$ diverges in the center. 
This solution is consequently not valid for small radii, where the
infall must deviate from the solution (\ref{Masswind}). Positive values
of $p$ correspond to outflows generated by the accretion flow and
reversing the argument for the total mass now taken away from the ADAF,
the solution is confined to small radii so that $\dot{M}_W$ is still
smaller than the mass accretion rate supplied at the outer radius.
The surface density of the flow implies a density $\rho \sim r^{-3/2+p}$,
which restricts the specific equation of state (\ref{EqState}) to those
with $\chi_\rho$,$\chi_T$ and $\Gamma_3$ being constants and the
temperature as a function of radius follows
%-------------------- 
\begin{equation} \label{TemC}
 T^{\chi_T} \sim r^{-1 +(1-\chi_\rho)(-3/2+p)}
\end{equation}  
%-----------------
so that the radial pressure gradient is independent of the thermodynamic
exponents and $P \sim r^{-5/2+p}$. The gas density in self-similar
solutions for advection-dominated accretion is unconstrained and
only the required inefficient cooling restricts the solutions to small
mass accretion rates below $\approx 10^{-2} \dot{M}_{\rm Edd}$
\citep{ny95b}. $\dot{M}_{\rm Edd}$ is the Eddington accretion rate
for an radiation efficiency of 10\% in terms of the rest mass of
the accreted material. 
%
%---------------------------------------------
\section{$\alpha$-ADAFs}\label{alphaD}
Whether the parametrisation of the wind, which is necessary
for the self-similar solution in \S\ref{SelfS}, is justified or not,
depends on boundary conditions, which the wind has to meet at large
distances, or the internal physics of outflows. While outflows might
naturally follow the self-similarity of ADAFs, the distribution of
stellar wind sources will determine, if a power-law dependence of the
infall rate is reasonable. The parametrisation of the viscosity
\citep{sun73} in equation (\ref{alphvis}) is based on dimensional
arguments and besides numerical simulations, which do not exist for
ADAF conditions, there is no way to determine what values for $\alpha$
are appropriate in the case of stationary ADAFs. Without information on
the strength of the magnetic fields from MHD-calculations, the adiabatic
exponent $\Gamma_3$ is also undetermined. We will discuss the
self-similar solutions as functions of $\alpha$ and $\Gamma_3$ as well as
for certain wind parameters and viscous force measures $\eta$ and $\eta_1$.
  %--------------------------------------
\begin{figure}[t]
  \plotone{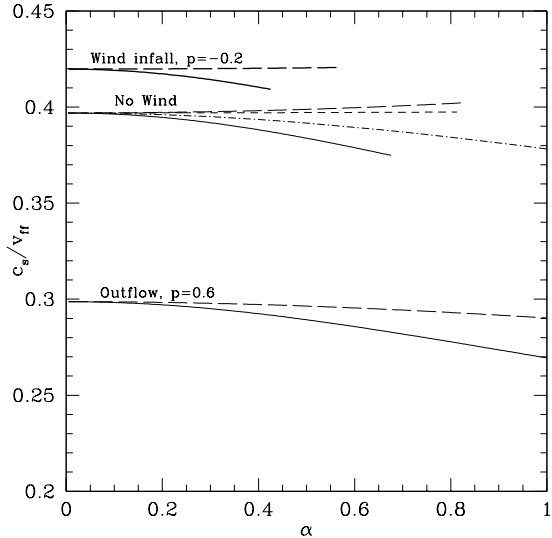} 
\figcaption[fig1.eps]{\label{CSalpha}
  Isothermal sound speed $c_s$ for ADAFs in units of the
  local free fall velocity $v_{\rm ff} = \sqrt{2GM/r}$. Both velocities 
  have the same radial scaling and the ratio is independent of radius.
  Solutions are shown for $\Gamma_3 = 1.4$ as a function of the
  viscosity parameter $\alpha$ as long as the corresponding $\Omega^2$
  is positive. The different sound speeds in all three
  cases of infall, no wind and outflow depend on the strength of 
  the radial viscous force. The solutions correspond to values of
  $\eta$ and $\eta_1$ given in Figure \protect\ref{veloalpha} 
  for accretion and rotation velocity. }
\end{figure}
%--------------------------------------

\begin{figure}
  \plotone{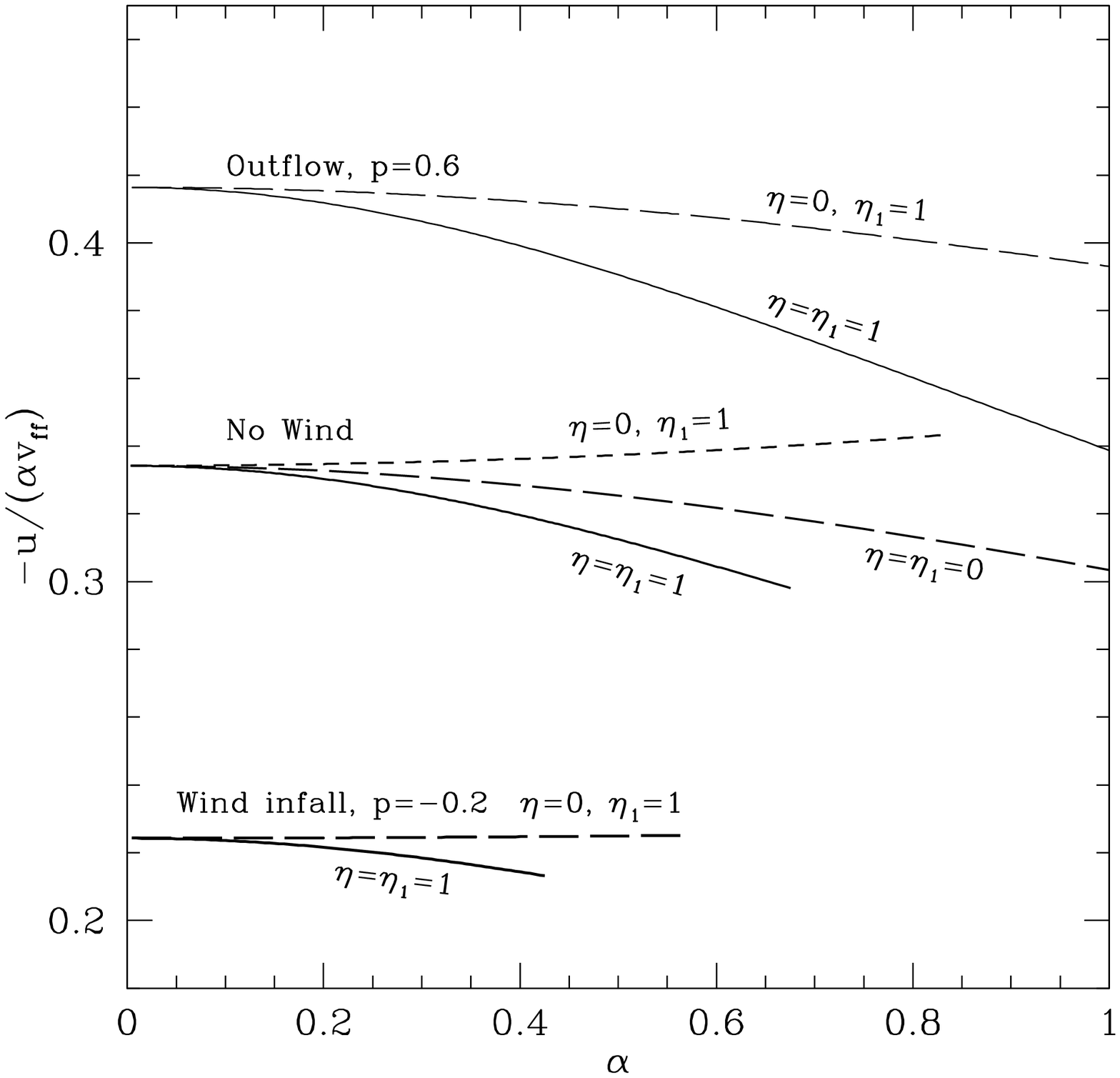}
  \plotone{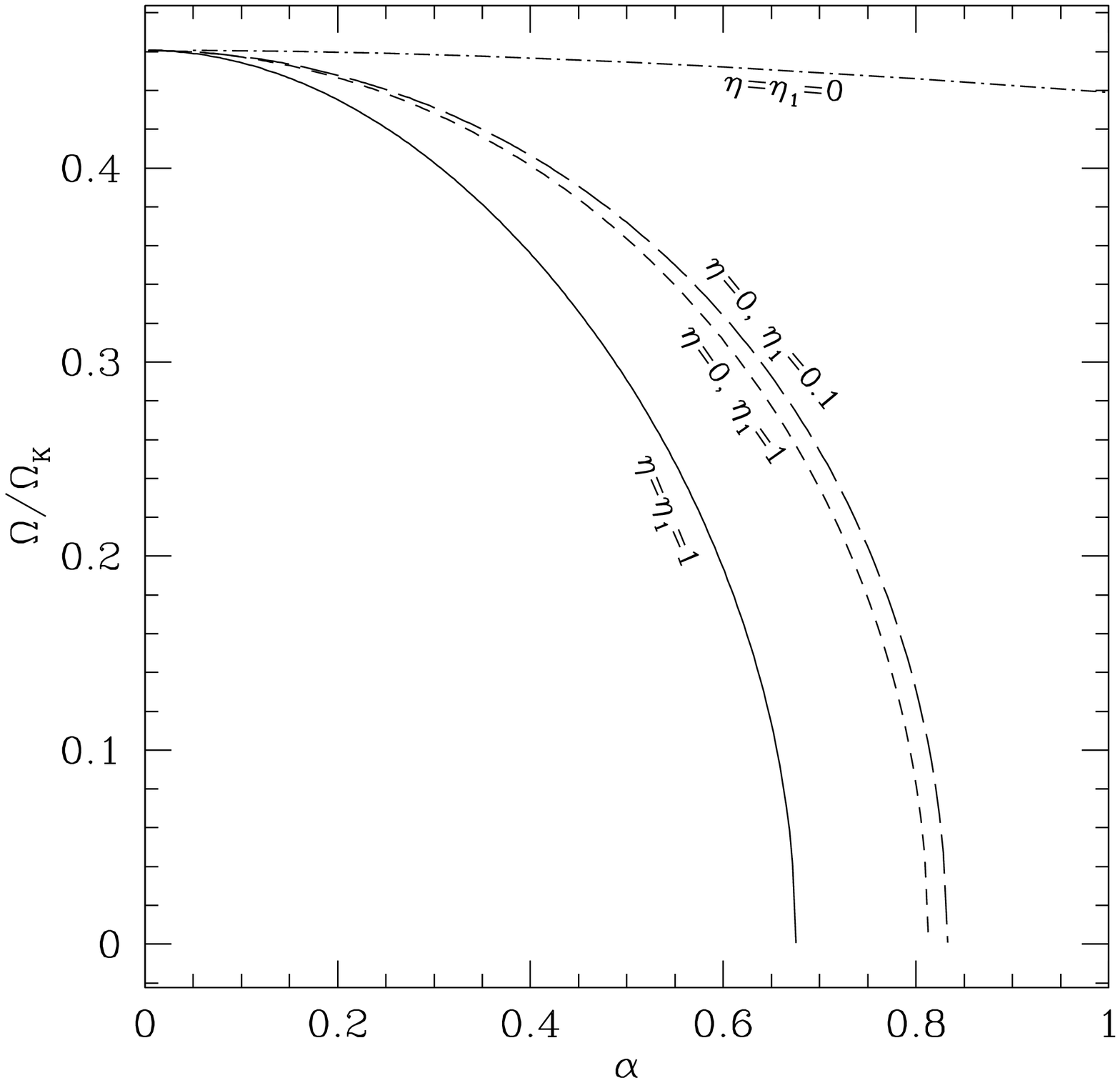} 
  \figcaption[fig2.eps]{\label{veloalpha}
  Radial accretion velocity $-u$ in units of $\alpha$ times the local
  free fall velocity (left) and angular rotation velocity $\Omega$ of
  the flow relative to the local Keplerian rotation (right).
  Solutions are given for $\Gamma_3 = 1.4$ and different strength
  of the radial viscous force. $\eta$ measures the strength of bulk
  viscosity and $\eta_1$ the $(rr)$ shear stress. Also shown is
  the standard ADAF solution without winds and no radial viscous
  force $\eta=\eta_1 =0$. The actual accretion velocity is almost linear
  in $\alpha$ and the difference at $\alpha \rightarrow 0$ is a
  difference in slope as a function of $\alpha$. The rotation
  velocity $\Omega$ is shown only for ADAFs without a wind.  }
\end{figure}

The self-similar ansatz in \S\ref{SelfS} reduces the dynamical
equations (\ref{Angu}), (\ref{AcVelo}) and (\ref{Temp}) to a set of
non-linear algebraic equations for the coefficients of sound speed $a$,
angular velocity $\Omega_0$ and accretion velocity $u_0$. The accretion
velocity is derived from equation (\ref{Angu}) in terms of sound speed
$a$ and $\alpha$
\begin{equation} \label{accv1}
  u_0 = \frac{3}{2} \alpha a^2 {\cal A} \qquad 
  {\cal A} = \frac{1+2p}{1+2p(1-\xi)}
\end{equation}
with a constant ${\cal A}$, which reflects the torque exerted by the
the wind on the flow. The combined wind parameter  ${\cal A}$ is 1 for 
a vanishing wind or a non-rotating wind. For rotating infall the
solution is restricted to moderate infall rates with $p> -1/2$. Otherwise
accretion would not be possible. For a given sound speed and accretion
velocity, the centrifugal barrier and therefore the angular velocity
follows from the radial momentum equation (\ref{AcVelo}), if we know the
viscous force measures. We find for the square of the angular velocity
\footnote{
\begin{equation} 
    {\cal F} = {\cal A}(1-2p(1-\xi_1))+\eta(5-2p)
    +\frac{4}{3}\eta_1(1+2p) \nonumber
\end{equation}
} 
%--------------- 
\begin{equation} \label{alphaOm}
   \Omega_0^2 = c^2 - \left(\frac{5}{2} - p\right) a^2 -
  \frac{9}{8}\alpha^2 {\cal A}\frac{a^4}{c^2} {\cal F}  \quad .
\end{equation}
%--------------
%\begin{equation} 
%    {\cal F} = {\cal A}(1-2p(1-\xi_1))+\eta(5-2p)
%    +\frac{4}{3}\eta_1(1+2p) \quad .
%\end{equation}
%--------------
We are left with a quadratic equation for the square of the sound speed.
One solution turns out to be irrelevant either by predicting $a^2$ to be
negative or leaving $\Omega_0^2$ in equation(\ref{alphaOm}) negative. 
But even the second root not always gives reasonable solutions, because
with increasing $\alpha$ and small changes in $a^2$ as seen in
Fig.\ref{CSalpha} and Fig.\ref{veloalpha}, $\Omega_0^2$ decreases and
becomes negative. Solutions for the accretion problem exist only for
values of the viscosity parameter smaller than a critical $\alpha_c$. 
The wind parameters for the solutions in Fig. \ref{CSalpha} and
\ref{veloalpha} have been chosen in a way that the infall or outflow
is in free fall or leaves with the escape speed. No radial momentum
of the wind is included and both infall and outflow are cold. While
the infall has no proper angular momentum, the outflow rotates with
the $\Omega$ of the accretion flow. With the requirements of a cold
outflow and total velocity of the wind being the escape speed, 
which both enter the energy equation (\ref{Temp}) in the same way, 
we make a minimal energy assumption for the extraction of internal energy
by the outflow. For the equation of state we take the natural
choice $\chi_\rho=1$ so that the mix of gas and magnetic field behaves
like an ideal gas. The ratio of specific heats equals $\Gamma_3$,
if we assume $\chi_T=1$. $\chi_T$ is not a parameter of our solutions,
but determines the actual temperature of the gas through equation
(\ref{TemC}). For energy equipartition we have $\Gamma_3 =1.4$
from equation (\ref{Ga3}) and we use this value if not stated otherwise.
For the cooling efficiency $1-f$ we assume a low rate of 1\%. For the
dynamics of the flow not the energy radiated away matters, but the energy
left in the flow. The actual cooling efficiency is unimportant as long
as it is small and $f\la 1$.
%---------------------------------------------------
%--------------------------------------
%--------------------------------------
\begin{figure}
  \epsscale{0.92}
  \plotone{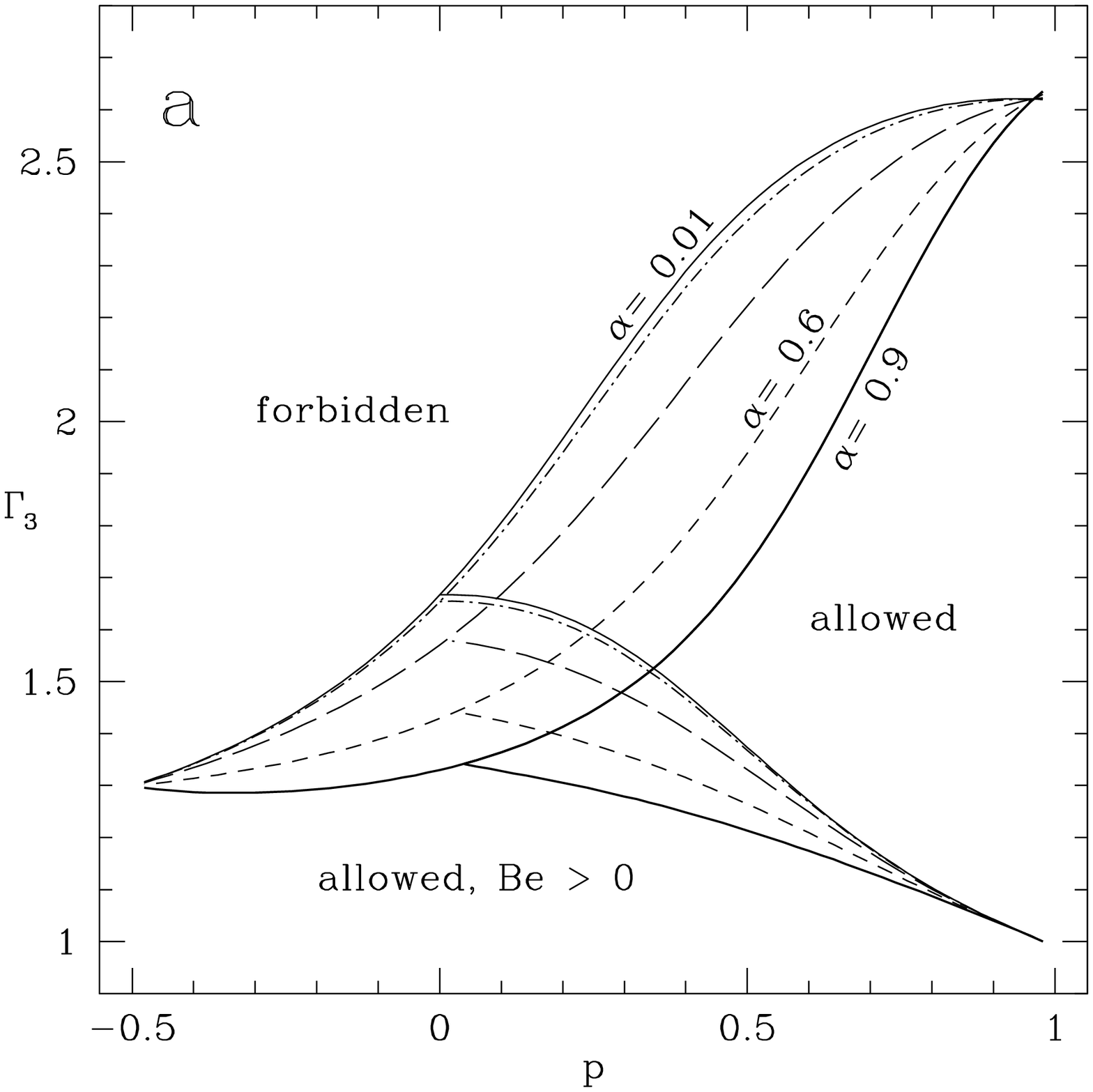}
  \plotone{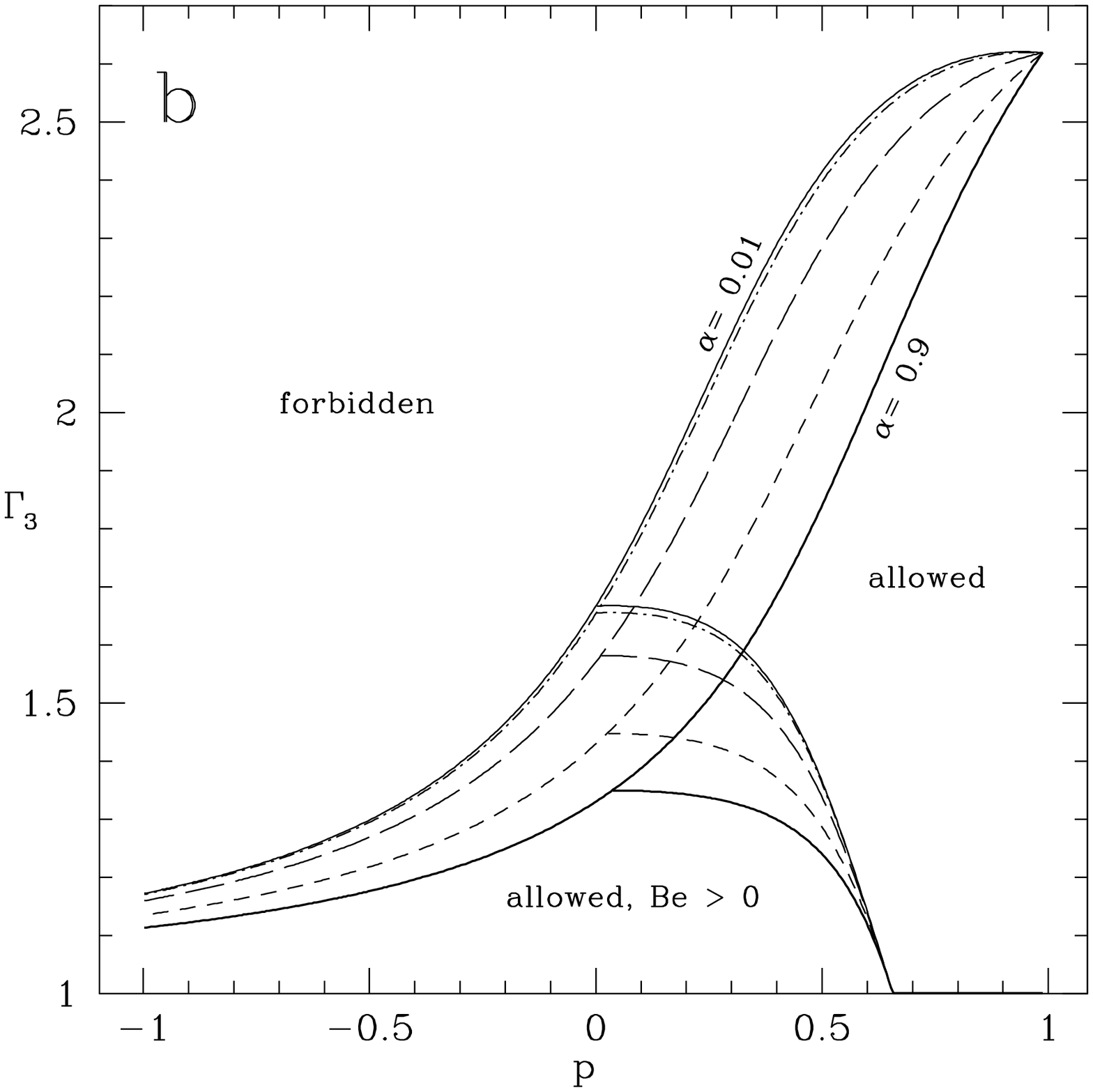}
  \figcaption[fig3a.eps,fig3b.eps]{ \label{GamPAl} 
   Excluded, allowed and marginally allowed regions of ADAF solutions
   with infall or outflows in the plane of wind parameter $p$
   and ratio of specific heats, which equals $\Gamma_3$ for 
   $\chi_\rho=\chi_T=1$ as mentioned in the text. $\Omega^2$
   becomes negative in the  forbidden region and we can distinguish 
   regions, where the Bernoulli number is either positive,
   representing unbound flows, or negative. Figure (a) shows the
   situation for corotation of wind and flow $\xi = 1$. In figure (b)
   a non rotating wind $\xi=0$ is assumed. The dividing line of
   allowed and forbidden regions depend on the strength of the radial
   viscous force, for which $\eta=0$ and $\eta_1=1$ is assumed.
   For increasing $\alpha$ the allowed region moves deeper into the
   outflow regime and dividing lines are shown for
   $\alpha = 0.01, 0.1, 0.3, 0.6, 0.9$.}
\end{figure}
%--------------------------------------

For an ADAF without winds it is possible to derive an analytic
solution for the sound speed and determine the maximum allowed $\alpha_c$ 
given by\footnote{
\begin{eqnarray} 
  {\cal B} & = & 2\left(5 \eta + \frac{4}{3}\eta_1\right)\left[\frac{4}{9} 
%  - (\Gamma_3-1)\left(\left(\frac{7}{3}-\Gamma_3\right)
%  \right.\right. \nonumber \\
%   & - & \left.\left. f\left(\frac{5}{3}-\Gamma_3\right)\right)\right] 
  - (\Gamma_3-1)
  \right. \nonumber \\
   & \times & \left. \left(\left(\frac{7}{3}-\Gamma_3\right)
   - f\left(\frac{5}{3}-\Gamma_3\right)\right)\right] 
   \nonumber \\ \label{balc}
   & + & \frac{32}{3}\eta_1 f (\Gamma-1)\left(\frac{5}{3}-\Gamma_3\right) 
\end{eqnarray}
}
%------------------------
\begin{equation}  \label{alpc}
  \alpha_c = \frac{\sqrt{8/9 - 2(\Gamma_3-1)(7/3 - \Gamma_3) + {\cal B}}}
  {f(\Gamma_3-1)(3\eta+4\eta_1)} \quad .
\end{equation}
%\begin{eqnarray} 
%  {\cal B} & = & 2\left(5 \eta + \frac{4}{3}\eta_1\right)\left[\frac{4}{9} 
%  - (\Gamma_3-1)
%  \right. \nonumber \\
%   & \times & \left. \left(\left(\frac{7}{3}-\Gamma_3\right)
%   - f\left(\frac{5}{3}-\Gamma_3\right)\right)\right] 
%   \nonumber \\ \label{balc}
%   & + & \frac{32}{3}\eta_1 f (\Gamma-1)\left(\frac{5}{3}-\Gamma_3\right) 
%  \quad .
%\end{eqnarray}
%------------------------
In that case ${\cal A}$ equals $1$ and the accretion velocity is 
$u_0 = (3/2)\,\alpha\;a^2 $. If the viscosity parameter is small,
the limit $\alpha \rightarrow 0$ provides a good approximation for
the sound speed 
%--------------
\begin{equation}
  a^2 \approx
  \frac{6f(\Gamma_3-1)}{3(\Gamma_3-1)(5f-2\chi_\rho)+10 - 6\chi_\rho}
\end{equation}
%--------------
and the angular velocity  $ \Omega = \sqrt{c^2 - (5/2)a^2}$ as seen in 
Fig.\ref{CSalpha} and Fig.\ref{veloalpha}. From equation (\ref{alpc})
and (\ref{balc}) we immediately see that  $1 < \Gamma_3 < 5/3$ 
is required for the no wind case and the familiar result arises that 
no solution exists for a non-relativistic ideal gas. It is also obvious
that the upper limit on $\alpha$ increases and becomes irrelevant
with vanishing radial viscous force. In other words, with increasing
viscous force measures $\eta,\eta_1$ ADAF solutions are restricted
to reasonably small values of $\alpha < \alpha_c$.

One additional criterion for the realization of accretion solutions
is the Bernoulli number
\begin{equation}
 {\rm Be} = -\frac{GM}{r} + r^2\Omega^2 + u^2 + e + \frac{P}{\rho}
\end{equation}
discussed by \citet{ny94,ny95} for ADAFs and \citet{blan99} for ADIOS.
While the total energy decreases inwards the Bernoulli number being the
total specific energy plus $P/\rho$ is positive and increases inwards
for ADAFs. This changes with outflows due to their cooling effect
by removing internal energy and reducing the Bernoulli number
of the remaining flow. Correspondingly an infall increases the Bernoulli
number of the flow. The combinations of allowed values for $\alpha$ 
and $\Gamma_3$ as a function of the wind strength $p$ is shown in 
Fig.\ref{GamPAl} for corotating ($\xi=1$) and non-rotating winds.
The statement that ADAFs exist for an ideal gas with negative
Bernoulli numbers, if an outflow is present \citep{blan99}, is confirmed,
even when the radial viscous force is included. The minimal possible
wind strength depends on the viscosity parameter $\alpha$ as
seen in Fig.\ref{GamPAl}.
%
%-------------------------------------------------------------------
\section{$\beta$-ADAFs with winds}\label{BetaDisk}
\begin{figure}
  \plotone{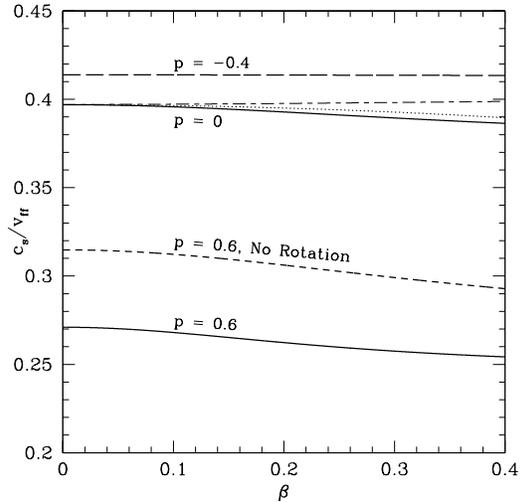}
  \plotone{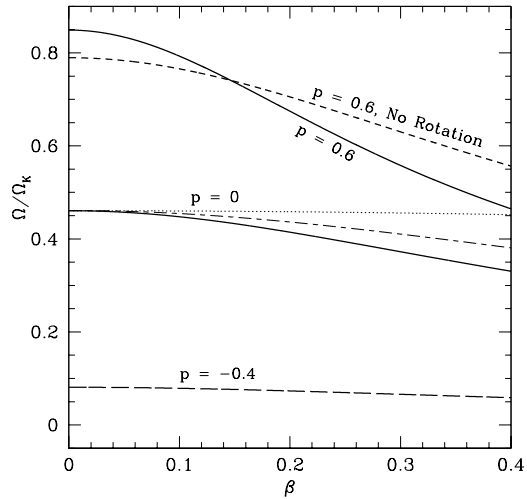}
  \figcaption[fig4a.eps,fig4b.eps]{\label{beta_F}
  Isothermal sound speed relative to the local free fall velocity
  (left) and angular velocity in units of the local Keplerian value
  (right) for $\beta$-ADAFs. For the no wind case $p=0$ the radial
  force measures are $\eta=\eta_1=1$ for the thick solid line,
  $\eta=0, \eta_1=1$ for the long-short dashed line and no radial
  viscous force $\eta=\eta_1=0$ for the dotted line. The infall
  solution $p=-0.4$ has no rotation of the infalling gas and for the
  outflow solution with $p=0.6$, the difference between a corotating
  outflow (thick solid line) and a non-rotating outflow is highlighted.}
\end{figure}

\begin{figure}
  \plotone{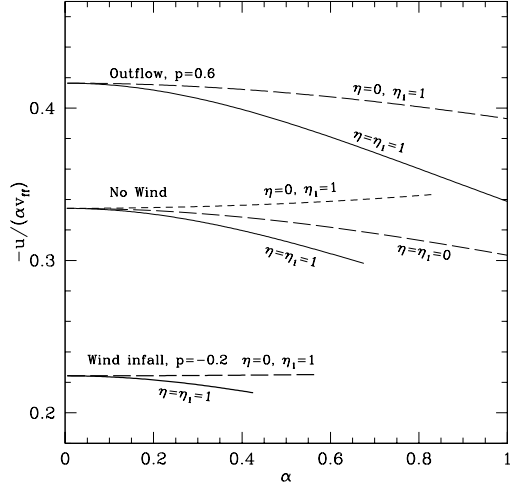}
  \plotone{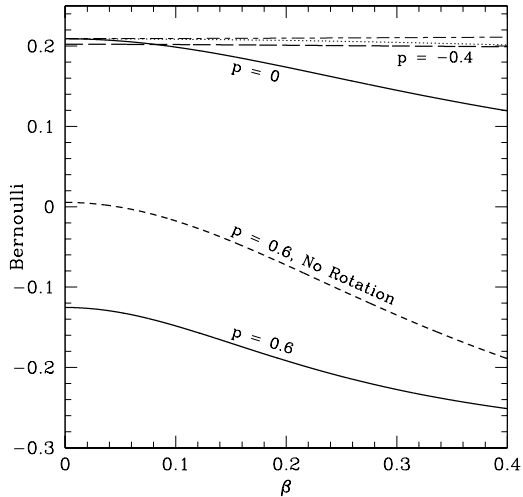}
  \figcaption[fig5a.eps,fig5b.eps]{\label{velbeta}
  Accretion velocity (left) and Bernoulli number (right) for 
  different $\beta$-ADAFs. The identification of solutions is the same 
  as in Fig. \protect\ref{beta_F}.}
\end{figure}

Following the arguments in \citet{dus20} the viscosity law should not
depend on the sound speed, if the actual velocity of turbulent eddies
is smaller than the sound speed in the gas. In the $\beta$-viscosity
law (\ref{BetaV}) the eddy velocity is determined from the actual
rotation velocity $v_{\rm eddy} \approx \sqrt{\beta}r\Omega$ with
the lessening $\sqrt{\beta}$ equally distributed between velocity
and length scale. The arguments given in \S\ref{Vis} suggest that
$\alpha$-viscosity can be seen as the shock-limited case of the
$\beta$-viscosity and the $\alpha$-ADAFs discussed in \S\ref{alphaD}
can be checked for consistency, if this hypothesis is correct. 
Specialising on the no wind case in the limit of small $\alpha$ we
find the ratio
%---------------
\begin{equation}
     \frac{c_s^2}{(r\Omega)^2} = \frac{f(\Gamma_3-1)}
      {(5/3-\Gamma_3\chi_\rho)}
\end{equation}
%---------------
and the $\beta$-viscosity can be applied for $\beta < 1.5 f$ with
the adaption of $\chi_\rho =1$ and $\Gamma_3 = 1.4$. Therefore the
$\beta$-viscosity is valid for ADAFs with low cooling efficiencies,
if the resulting flows are similar to $\alpha$-disks. The scale length
of the $\alpha$-viscosity is the vertical scale height $H =
c_s/\Omega_K$, which turns out be $H \approx 1.55 r \sqrt{f/(1.6+6f)}
\approx 0.6 r$ for $\alpha$-ADAFs in the no wind case and the limit
$\alpha\rightarrow 0$. The length scale of the $\beta$-viscosity
is $\sqrt{\beta}r$ and we expect the same sound speeds and angular
velocities for $\alpha$- and $\beta$-ADAFs below $\beta \la 0.3$.
This is verified by comparing Fig.\ref{CSalpha} and \ref{veloalpha}
with Fig.\ref{beta_F}. The actual solution for $\beta$-ADAFs is
derived in the same way as for $\alpha$-ADAFs in \S\ref{alphaD} and 
we find the accretion velocity depending explicitly on $\Omega$ and
not $c_s^2$.  
%---------------   
\begin{equation} \label{velob}
  u_0 =-\frac{3}{2}\beta\Omega_0{\cal A}
\end{equation}  
%--------------
The constant ${\cal A}$ is defined in equation (\ref{accv1}). The
increased accretion velocity seen in Fig.\ref{velbeta} compared to 
$\alpha$-ADAFs is explained by the larger value of $\Omega_0$ in
equation (\ref{accv1}) relative to the sound speed squared in case
of the $\alpha$-viscosity. The angular velocity of the accretion flow
can be given\footnote{\[ {\cal H}
  = \left((5-2p)\eta+4/3(1+2p)\eta_1\right){\cal A}\]}
%----------------
\begin{equation}
  \Omega_0^2 = \frac{1-(5/2-p)a^2}{9/8\left[(1-2p(1-\xi_1)){\cal A}^2
  +{\cal H}\right] \beta^2+1} 
\end{equation}  
%\begin{equation}
%  {\cal H} = \left((5-2p)\eta+4/3(1+2p)\eta_1\right){\cal A}
%\end{equation}  
%--------------
including the reaction of the flow due to the presence of winds. 
Turning to the energy equation shows that two obvious solutions
exist for the sound speed. First an non-rotating and non-accreting
solution 
\begin{equation}
  a^2 = \frac{2c^2}{5-2p} \quad ,
\end{equation}
where the gravitational attraction is completely balanced by the pressure 
gradient. The gas is in hydrostatic equilibrium and constitutes an 
optically thin atmosphere.
%--------------------------------------
%--------------------------------------
%--------------------------------------
%--------------------------------------
This is not a viable solution for a
black hole, because general relativity does not allow a pressure
supported atmosphere close to the horizon. The second solution is
an accreting and rotating flow as shown in Fig.\ref{beta_F}. The gas is
cooler than in the atmosphere solution and strong outflows in corotation
with the flow or even non-rotating outflows reduce the sound speed
further so that the accreted gas can be gravitationally bound to the
central mass.  For a corotating, cold outflow, which reaches infinity
with the escape velocity from the radius of origin $\xi_2 =2$, and
a simple equation of state $\chi_\rho = 1$ we can give the sound speed
for a case of negligible radial viscosity $\eta = \eta_1 = 0$:
\footnote{\begin{eqnarray*}
  \Xi & = & \frac{{\cal A}}{24}\left[4(6-p)(1-2p)(1+\beta^2{\cal A}^2)
 \right.  \\ & + & \left.
 \beta^2{\cal A}^2(3(1+4p)^2-p(17+2p))\right] 
  \\
  & + & \frac{{\cal A}(1-p)}{\Gamma_3-1}\left(\frac{2}{3}
  +\frac{3}{4}\beta^2{\cal A}^2(1-2p)\right)
\end{eqnarray*}
}
\begin{equation}
  a^2 = \frac{f-\left[3(3/4-p)\beta^2{\cal A}^2+1\right]p{\cal A}}
  {(5/2-p)f-\Xi} \quad .
\end{equation}
%\begin{eqnarray}
%  \Xi & = & \frac{{\cal A}}{24}\left[4(6-p)(1-2p)(1+\beta^2{\cal A}^2)
% \right. \nonumber \\ & + & \left.
% \beta^2{\cal A}^2(3(1+4p)^2-p(17+2p))\right] 
%  \nonumber \\
%  & + & \frac{{\cal A}(1-p)}{\Gamma_3-1}\left(\frac{2}{3}
%  +\frac{3}{4}\beta^2{\cal A}^2(1-2p)\right) \quad .
%\end{eqnarray}
For non-rotating and cold infall, which is in free-fall for $p<0$,
or outflows barely reaching infinity ($\xi_2 = 1$) with $p<0.36$ 
the ratio of specific heats has to be smaller than a limiting value 
\begin{equation}
 \Gamma_3-1 <  2\frac{1-p}{3-9p+2p^2} 
\end{equation}
at which the rotation of the flow stops and no solutions exist for 
$\Gamma_3$ larger than that. This limit for $\Gamma_3$ equals $5/3$ 
for ADAFs without winds or outflows and is the well known result
that ADAFs do not exits for an ideal equation of state.
The consistency of $\beta$-ADAFs follow from the above predictions
based on $\alpha$-flows. With infall into the flow the sound speed
increases and the rotation velocity decreases. The $\beta$-viscosity
law is therefore more likely to be applicable to ADAFs with infall
than with outflows. 
%
%---------------------------------------------------------------------
\section{ADAFs with shear-limited viscosity and winds}\label{shearADAF}
\begin{figure}
  \epsscale{0.92}
  \plotone{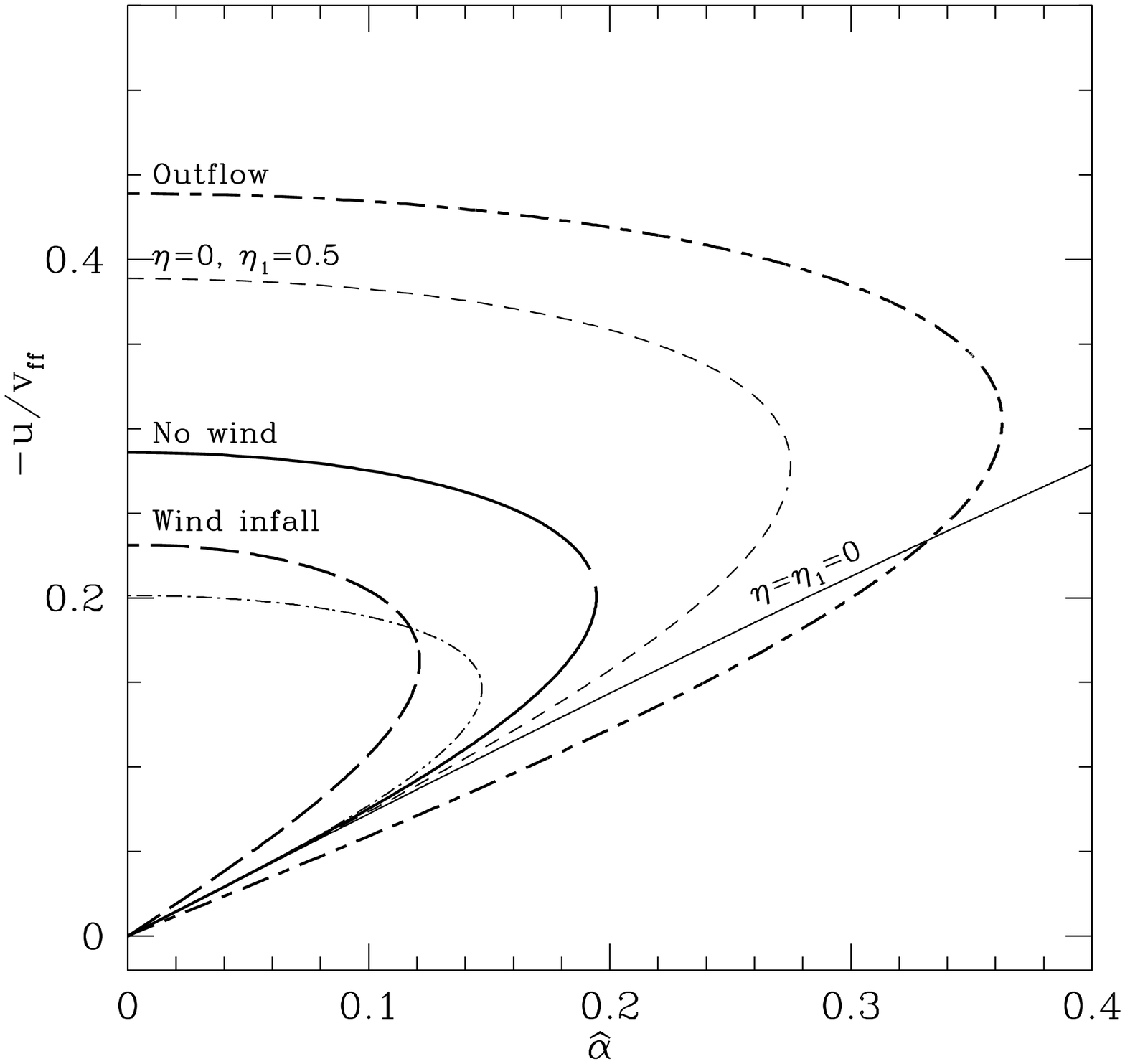}
  \plotone{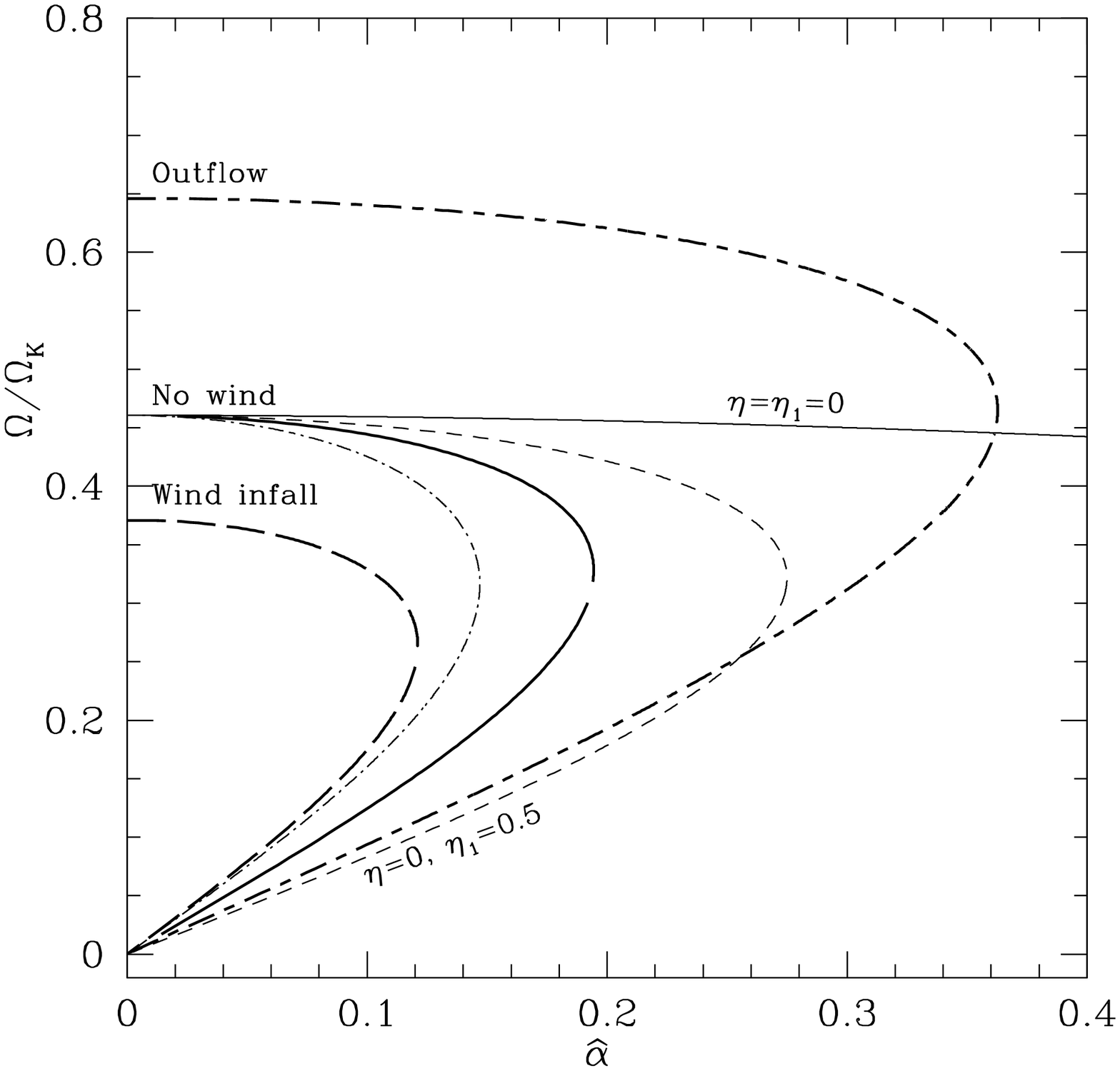}
  \figcaption[fig6a.eps,fig6b.eps]{\label{modalVeloO}
  Accretion velocity (left) and rotation (right) for ADAFs with the 
  shear-limited viscosity law. The ratio of specific heats equals
  $\Gamma_3 = 1.4$ and wind parameters are $p=0.2$ for the corotating
  outflow and $p=-0.1$ for infall with no angular momentum of the
  infalling material. The solution without radial viscous force is
  plotted as a solid thin line available for all values of
  $\hat{\alpha}$. Solutions with different viscous force measures 
  are shown for the no wind case $p=0$ with $\eta=0, \eta_1=1$ as the 
  thick solid line,  $\eta=1, \eta_1=1$ the thin dash-dotted line and 
  $\eta=0, \eta_1=0.5$ the short dashed line as indicated. For each
  solution the branch with large radial velocity shows slower rotation
  and vice versa.  
 }
\end{figure}

\begin{figure}
  \plotone{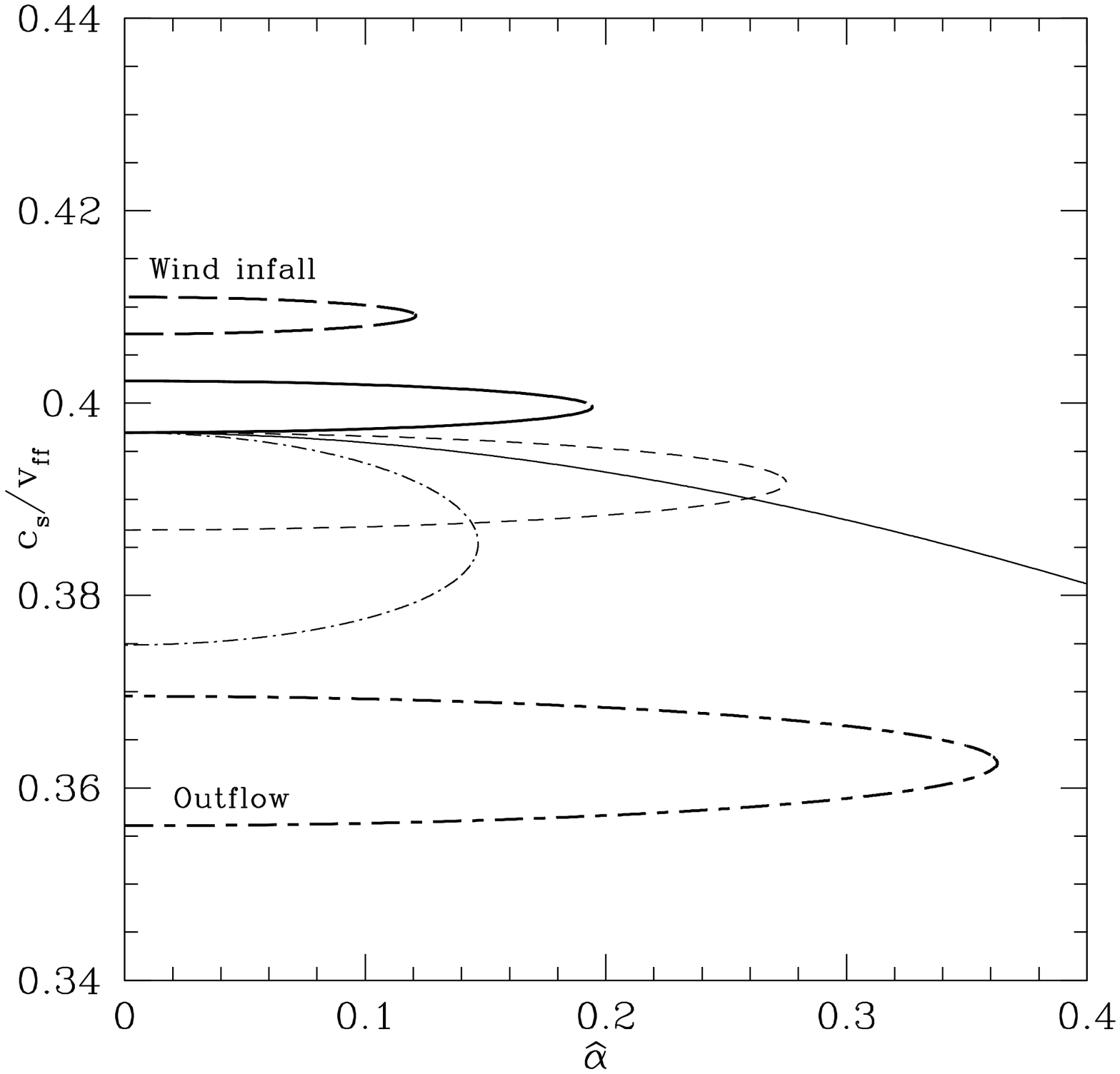}
  \plotone{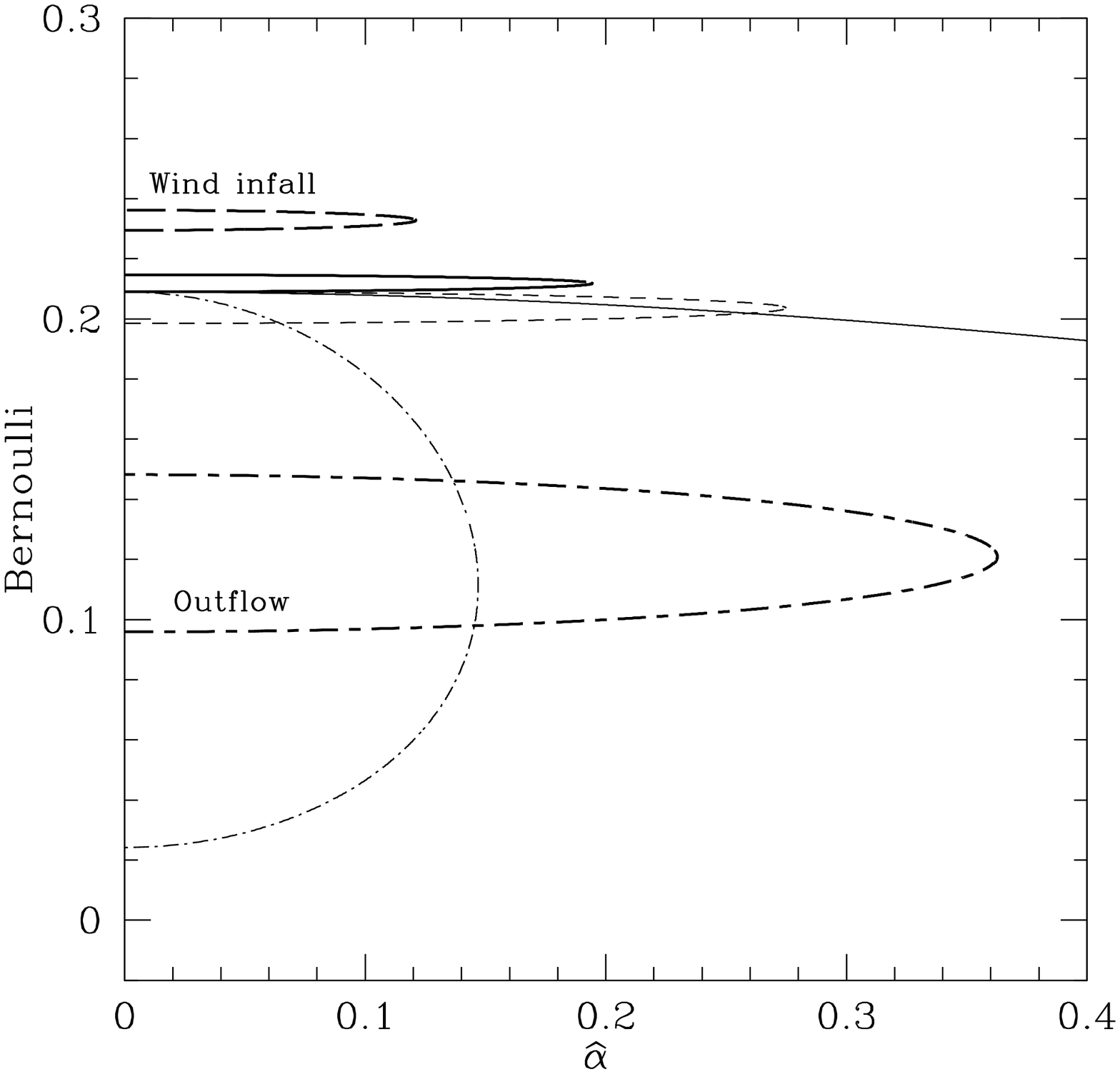}
  \figcaption[fig7a.eps,fig7b.eps]{\label{soundba}
  Isothermal sound speed and Bernoulli number for the two branch ADAF
  model with shear-limited viscosity law. The identification of
  solutions is by line type and is the same as in Fig.
  \protect\ref{modalVeloO}. The branch of each solution with large
  sound speed corresponds also to large accretion velocities and
  slow rotation. The Bernoulli number for outflow solutions is
  smaller for the rapidly rotating branch with small sound speed
  compared to the slowly rotating branch. In all other cases
  the Bernoulli number of the slowly rotating solution is smaller.  }
\end{figure}

\begin{figure}[t]
  \plotone{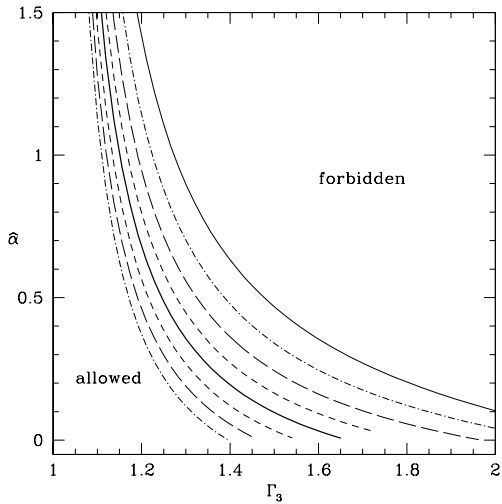}
  \figcaption[fig8.eps]{\label{maxalc}
  Dividing lines between regions with and without solutions for ADAFs
  in the $\hat{\alpha}$-$\Gamma_3$ plane of viscosity parameter and
  ratio of specific heats for $\chi_\rho =\chi_T = 1$ and only $(rr)$
  shear stress considered for the radial viscous force $\eta=0$,
  $\eta_1=1$. The regions of allowed and excluded combinations of
  $(\hat{\alpha},\Gamma_3)$ depend on the wind parameter $p$, which
  varies from $p = -0.3$ to $p=0.4$ in steps of $\Delta p =0.1$.
  The dividing line moves to larger $\hat{\alpha}$ and $\Gamma_3$ for
  increasing wind parameter $p$. The no wind case $p=0$ is emphasised
  as the thick solid line. }
\end{figure}

The motivation for a shear limited viscosity law is the observation 
that the standard $\alpha$-viscosity agrees with the general 
$\beta$-viscosity law only for Keplerian accretion disks with a
shock-limited eddy velocity. In the shear-limited viscosity 
%---------------
\begin{equation}\label{shearvis}
  \nu = \hat{\alpha}\frac{c_s^2}{\Omega}
\end{equation}
%-------------
we assume that the eddy 
velocity is indeed shock-limited and the mean free path of eddies is
determined by the radial distance of differentially rotating shells,
which allows the exchange of eddies along radial paths in a comoving
frame of one shell. If the eddy velocity is the sound speed, the
maximal distance of shells follows from a Taylor expansion of
$\Omega$ in $\Delta v_\phi$ 
%--------------
\begin{equation}
  \Delta r \la \frac{c_s}{2\Omega} \quad .
\end{equation}
%--------------
The length scale over which shells can interact is larger for
sub-keplerian rotation. This is the main difference of equation
(\ref{shearvis}) to the original $\alpha$-viscosity. The consequence
for the accretion velocity 
%--------------
\begin{equation} 
  u_0 = \frac{3}{2} \hat{\alpha} \frac{a^2}{\Omega_0} {\cal A} 
\end{equation}
%--------------
is an explicit dependence on the inverse of $\Omega_0$ contrary
to $\beta$-ADAFs in equation (\ref{velob}). The combined wind parameter
${\cal A}$ is defined in equation (\ref{accv1}). The square of the
angular velocity follows  from a quadratic equation with two roots
\footnote{
\begin{eqnarray*}  
 \Psi & = & 
   \frac{9}{8}\hat{\alpha}^2{\cal A}\left[(1-2p(1-\xi_1)){\cal A} \right.  \\ 
  & + & \left. \frac{4}{3}\eta_1(1+2p)+\eta(5-2p)\right] 
\end{eqnarray*}
}
%-------------- 
\begin{equation} \label{twobOmega}
   \Omega_0^2 = \frac{1}{2} - \frac{5-2p}{4}a^2 
   \pm \sqrt{\left(\frac{1}{2} - \frac{5-2p}{4}a^2\right)^2
   - a^4 \Psi}  .
\end{equation}
%--------------
%--------------
%\begin{eqnarray}  
% \Psi & = & 
%   \frac{9}{8}\hat{\alpha}^2{\cal A}\left[(1-2p(1-\xi_1)){\cal A}
%  \right. \nonumber \\ 
%  & + & \left. \frac{4}{3}\eta_1(1+2p)+\eta(5-2p)\right]  \quad .
%\end{eqnarray}
%--------------
In the limit of vanishing viscosity parameter $\hat{\alpha}$ the angular 
velocity has two obvious solutions 
%--------------
\begin{equation} 
 \hat{\alpha} \rightarrow 0 \quad \Rightarrow \Omega_0 \rightarrow \left\{
\begin{array}{l} \sqrt{1 - (5/2-p)a^2 } \\ {\rm or} \\
                  0
    \end{array}\right. \quad .
\end{equation}
%--------------
For small $\hat{\alpha}$ the rotation for the second solution is
proportional to $\hat{\alpha}$ and positive. The existence of these
slowly or non-rotating ADAFs depends on non-vanishing force measures
$\eta, \eta_1 \ne 0$. The solution disappears without a radial viscous
force. The viscosity in non-rotating ADAFs tends to a finite value
at $\hat{\alpha}=0$ 
%---------------
\begin{equation}
  \nu = a\ \sqrt{\frac{1 - (5/2-p)a^2}{a^2 \Psi}}\ c\ \sqrt{ s r_G} 
\end{equation}
%--------------
and so does the accretion velocity. If we interpret
$\hat{\alpha} c_s/\Omega$ as the viscous length scale, it does not
diverge for small $\hat{\alpha}$. The sound speed also tends to a
positive value as seen in the upper solution branches for the sound
speed in Fig.\ref{soundba} and accretion velocity in Fig.\ref{modalVeloO}
corresponding to the lower branches for the angular velocity
in Fig.\ref{modalVeloO}. The energy equation is itself quadratic
in $a^2$ for either choices of $\Omega_0$ in equation (\ref{twobOmega}). 
Only one solution satisfies the condition
%----------------
\begin{equation}
  a^2 \le \frac{2}{5 - 2p}
\end{equation}
%----------------
for real values of $\Omega$. We end up with two solution branches, 
which join at a critical $\hat{\alpha}_c$.
%\onecolumn
%--------------------------------------
%--------------------------------------
%--------------------------------------
%--------------------------------------
%\twocolumn
Beyond  $\hat{\alpha}_c$ no 
physical solution exists for the accretion problem. Besides the slowly
rotating solution discussed above, we find more familiar ADAFs,
where the accretion velocity is linear in $\hat{\alpha}$ as seen in
Fig.\ref{modalVeloO}. For small viscosity parameters, the solution is 
independent of the force measures $\eta,\eta_1$. Real solutions for the
square of the sound speed in the no wind case are only possible for
%--------------
\begin{equation}
  \hat{\alpha} \le \frac{(5/3-\chi_\rho\Gamma_3)}
  {3f(\Gamma_3-1)\sqrt{\eta+(4/3)\eta_1}} \quad .
\end{equation}
%--------------
The situation does not differ from the standard $\alpha$-viscosity 
in so far as no upper limit for $\hat{\alpha}$ exists in the limit
of vanishing radial viscosity $\eta=\eta_1 =0$ in the radial momentum
equation (\ref{AcVelo}) as long as the ratio of specific heats
is in the range $1 < \Gamma_3  \le 5/3$. In the limit of vanishing
viscosity $\hat{\alpha}\rightarrow 0$ the rotating solution coincides
with the unique ADAF solution without winds and radial viscosity.
For the simplest equation of state $\chi_\rho = 1$ the sound speed
in the limit $\hat{\alpha} \rightarrow 0$ is
%--------------
\begin{equation}
  a^2 = \frac{2\epsilon f}{2\epsilon^2 +5 \epsilon f 
%+ \left((3/2)  \hat{\alpha} f\right)^2
  } \quad , 
  \qquad \epsilon = \frac{5/3-\Gamma_3}{\Gamma_3-1}\quad .
\end{equation}
%--------------
The critical $\hat{\alpha}_c$, which sets the upper limit for the
existence of ADAFs with shear-limited viscosity, depends not only
on the force measures $\eta, \eta_1$, but also on the presence of winds.
The same pattern as for the standard $\alpha$-viscosity appears,
in which solutions with outflows are possible for a wider range
of $\hat{\alpha}$ than for infall solutions as seen in
Fig.\ref{modalVeloO} and Fig.\ref{soundba}. It turns out that the
slowly rotating branch of solutions produces smaller Bernoulli numbers
than the fast rotators for infall and the no wind case $p=0$. Only for
outflow solutions is the fast rotator preferred with smaller
Bernoulli numbers. But in all cases shown in Fig.\ref{soundba},
the Bernoulli number is positive and no gravitationally bound flow
is found. This changes if more energy is extracted by the wind and the
fast rotating outflow solution shown in the figures has negative
Bernoulli numbers, if the outflow is cold and reaches a terminal velocity
equal to the escape speed at its origin. The allowed combinations
of $\hat{\alpha}$ and $\Gamma_3$ regardless of Bernoulli numbers
are shown in Fig.\ref{maxalc}. The dividing lines in the $(\hat{\alpha},
\Gamma_3)$-plane are vertical without a radial viscous force and
bend in the way drawn in the figure, when the $(rr)$ component of
the stress tensor is included. 
%--------------------------------------
%--------------------------------------
%
%----------------------------------- 
\section{Conclusions}\label{Conclud} 
We have investigated the importance of a radial viscous braking force
for advection-dominated accretion flows (ADAFs) in the presence of
infall or outflows. Under the assumption of almost isotropic turbulence,
bulk viscosity and $(rr)$ component of the viscous stress tensor provide
an efficient brake of the rapid radial accretion of hot gas in ADAFs
and produce additional heat due to viscous friction. This opens a
second channel for transfer of kinetic into internal energy and
supports in part the radial motion so that the shear due to
differential rotation and the centrifugal barrier is reduced in ADAFs
with radial viscous braking.

We derived self-similar solutions of ADAFs for three different viscosity
laws. The standard $\alpha$-viscosity produces solutions, which show
more and more sub-keplerian rotation with increasing $\alpha$. The
solutions terminate at a critical $\alpha_c$, which depends strongly on
the ratio of specific heats so that the greatest possible $\alpha$ tends
to zero for a non-relativistic ideal gas. At the critical $\alpha_c$
the rotation of the flow vanishes and a purely radial inflow appears,
which is not only supported by a pressure gradient, but also by the viscous
braking force, provided the flow is still turbulent in the absence of
differential rotation. This limit is reminiscent of Bondi accretion 
\citep{bond52} in the presence of effective turbulent viscosity. 
The transition from ADAFs to Bondi accretion nonetheless affords
a suspicious fine tuning in $\alpha$.

The $\beta$-viscosity law of \citet{dus20} based on geometrical
arguments in the absence of shock-limited turbulence allow ADAF
solutions for all reasonable values of $\beta$. The estimates of $\beta$
as inverse of the critical Reynolds number of the flow suggests
small $\beta$s, for which the solutions do not differ from
$\alpha$-ADAFs. No transition to a Bondi like flow is possible in this
case and no upper bound on $\beta$ exists.

We showed that it is possible to derive a shear-limited viscosity law
from the $\beta$-viscosity mentioned above, for which the transition
to non-rotating Bondi like accretion with turbulent viscosity occurs
naturally. The now familiar ADAF solutions with sub-keplerian rotation
at $\hat{\alpha} \rightarrow 0$ join with a second branch of solutions
at a maximal sustainable $\hat{\alpha}$. Similar to the $\alpha$-ADAFs
no accretion flow with larger $\hat{\alpha}$ are possible. The second
solution branch is a hot, slowly rotating, and rapidly accreting solution
even for small values of $\hat{\alpha}$. The transition to viscous
Bondi accretion occurs from the slowly rotating to the non-rotating
solution in the limit $\hat{\alpha} \rightarrow 0$, where $\Omega$
is linear in $\hat{\alpha}$. A finite viscous force is present in
these flows even in the limit $\hat{\alpha}=0$. In all cases the
connection of ADAFs to non-rotating flows depends on the presence
of a radial viscous force.

The observation that ADAFs generally possess positive Bernoulli numbers
lead to the idea \citep{ny95, blan99} that ADAFs are good candidates
for the production of outflows. In that way the accretion flow loses
energy to the outflow and the remaining material is left with negative
Bernoulli numbers and gravitationally bound to the central accreting
mass. We confirm that statement in the presence of a radial viscous
force for $\alpha$- and $\beta$-ADAFs and show the back-reaction of
outflows on ADAFs for different outflow characteristics. Most noticeably
is the cooling effect and the increased accretion velocity of the
remaining ADAF as angular momentum and internal energy is carried away.
Outflows with a minimal energy assumption for the extracted energy have
to be fairly massive $p\ga 0.35$ to lower the Bernoulli number and
leave a bound flow in case of the shear-limited viscosity. It is much
easier to get a bound flow, if the minimal energy assumption is violated
and the terminal velocity of the outflow is the escape speed from the
origin of the out-flowing material.

If  a natural choice of the viscosity parameter---either $\alpha$ or 
$\hat{\alpha}$---exists and provided the strength of the radial viscous
force can be estimated from isotropic turbulence, then outflows
are inevitable for certain equations of state with ratios for specific
heats close to $5/3$. This conclusion is independent of arguments
based on the positiveness of Bernoulli numbers for ADAFs.

The scenario of thin disk evaporation in binary systems \citep{liu99}
or the transition from cooling flows to ADAFs in low-luminosity cores
of elliptical galaxies \citep{qua99b} can explain the existence
of ADAFs in these systems. The formation of an ADAF, which is the
most promising model for the spectral energy distribution of
Sgr A$^*$ in the Galactic Center, cannot proceed in either way.
We suggest that an ADAF in the Galactic Center forms out of stellar
wind infall \citep{cok99} and the transfer of kinetic energy of the
infall into internal energy of an advection-dominated flow. The even
larger Bernoulli numbers produced in this way wound give rise to
subsequent outflows and reduce the mass accretion rate inferred from
infall calculations by \citet{cok99} to the smaller accretion rates
predicted from spectral fitting \citep{qua99} of ADAF models to
Sgr A$^*$. 

{\acknowledgments
I thank Ramesh Narayan and Wolfgang Duschl for helpful conversations
and challenging discussions. This work has been supported through
DAAD fellowship D/98/27005. }

\appendix
\twocolumn
\section{The radial viscous force} \label{VisF}
The radial viscous force in spherical coordinates is derived
from the tensor divergence of the shear stress tensor $t_{ij}$ and the
bulk viscous force from the divergence of bulk viscosity and compression.
The stress tensor is assumed to be proportional to the shear tensor
${\cal D}$  which  is defined as
\begin{equation}
  {\cal D}_{ij} = \frac{1}{2}(v_{i;j} + v_{j;i}) - \frac{1}{3}
  g_{ij}v^l_{;l}
\end{equation} 
with the kinematic viscosity $\mu = \nu\rho$ as a scalar function
relating shear and stress $t_{ij} = \mu {\cal D}_{ij}$. Here $v_i$ are
velocity vector components, $g_{ij}$ the metric tensor
and $v_{i;j}$ implies covariant differentiation of the velocity. 
The radial viscous force derived from the shear tensor is 
\begin{eqnarray}
  f_\nu & = & (2 \mu {\cal D}^{r j})_{;j}  \\ & = & \nonumber
  \frac{4}{3}\left(\frac{\partial}{\partial r}
  \left[\mu r \frac{\partial}{\partial r}\left(\frac{u}{r}\right)\right]
  +3\mu\frac{\partial}{\partial r}\left(\frac{u}{r}\right)\right) \quad ,
\end{eqnarray}  
where axial-symmetry and $v_\theta = 0$ is assumed and $u$ is the radial 
velocity as used throughout the paper. The corresponding heating rate
due to this viscous friction is
\begin{equation} \label{heatrr}
 q^+_{rr} = 
  \frac{4}{3}\mu r^2\left(\frac{\partial}{\partial r}
  \left(\frac{u}{r}\right)\right)^2 \quad .
\end{equation} 
The contribution from compression and bulk viscosity $\zeta$ to the
radial viscous force is 
\begin{equation}
  f_{\rm bulk} = (\zeta v^l_{;l})_{;r} = \frac{\partial}{\partial r}
    \left(\frac{\zeta}{r^2}\frac{\partial}{\partial r}(u r^2)\right)
\end{equation} 
and the corresponding heating rate is
\begin{equation} \label{heatbulk}
 q^+_{\rm bulk} = \left(\frac{\zeta}{r^2}\frac{\partial}{\partial r}
  (u r^2)\right)^2 \quad .
\end{equation} 
Multiplying by $r$ and performing polar integration of the force gives
the radial viscous force used in (\ref{rrshear}) and (\ref{Fbulk})
\begin{equation}
  F_{rr} = \frac{4r}{3}\left(\frac{\partial}{\partial r}
  \left[\nu \Sigma \frac{\partial}{\partial r}\left(\frac{u}{r}\right)
  \right]+3 \nu \Sigma
  \frac{\partial}{\partial r}\left(\frac{u}{r}\right)\right)
\end{equation}
and  
\begin{equation}
  F_{\rm bulk} = 
    r\frac{\partial}{\partial r}
    \left(\frac{\nu \Sigma}{r^3}\frac{\partial}{\partial r}(u r^2)\right)
\end{equation} 
with the replacement of the bulk viscosity by the effective turbulent
viscosity $\nu$. The heating rates used in equation (\ref{heatQ})
are derived from equation (\ref{heatrr}) and (\ref{heatbulk}) in the
same way. The force measures $\eta,\eta_1$ used in the text have
been omitted here.

\end{document}